\documentclass[prd,nofootinbib,showpacs,12pt]{revtex4}

\usepackage{graphicx}

\begin{document}




%
%

\title{Predictions for high-energy $pp$ and $\bar pp$ scattering \\ from a
finite sum of gluon ladders}

\author{R.~Fiore\,$^a$}\email{fiore@cs.infn.it}
\author{L.~Jenkovszky\,$^b$}\email{jenk@bitp.kiev.ua}
\author{E.~Kuraev\,$^c$}\email{kuraev@thsun1.jinr.ru}
\author{A.~Lengyel\,$^d$}\email{alexander-lengyel@rambler.ru}
\author{Z.~Tarics\,$^d$}\email{iep@iep.uzhgorod.ua}
\affiliation{$^a$Dipartimento di Fisica, Universit\`{a} della
Calabria and INFN, Gruppo collegato di Cosenza
 \\
I-87036 Arcavacata di Rende, Cosenza, Italy \\
$^b$Bogolubov Institute for Theoretical Physics, National Academy of Sciences of Ukraine, Kiev, 03680 Ukraine\\
$^c$Bogolubov Laboratory of Theoretical Physics, Joint Institute for Nuclear Research, Dubna, 140980 Russia\\
$^d$Institute of Electron Physics, National Academy of Sciences of
Ukraine, Uzhgorod, 88017 Ukraine}
\begin{abstract}
An eikonalized elastic proton-proton and proton-antiproton
scattering amplitude $F(s,t)$, calculated from QCD as a finite sum
of gluon ladders, is compared with the existing experimental data
on the total cross section and the ratio $\rho(s,0)=Re F(s,0)/Im
F(s,0)$ of the real part to the imaginary part of the forward
amplitude. Predictions for the expected LHC energies are given.
\end{abstract}


\maketitle

\section{Introduction}
According to the common belief, the Pomeron in QCD corresponds to
an infinite sum of gluon ladders with Reggeized gluons,
resulting~\cite{FKL,BL,L} in the so-called supercritical behavior
$\sigma _t\sim s^{\alpha_P (0)-1}$, where $\alpha_P (0) >1$ is the
intercept of the Pomeron trajectory, as discussed in Ref.
\cite{BFKLP}. In that approach, the main contribution to the
inelastic amplitude and to the absorptive part of the elastic
amplitude in the forward direction arises from the multi-Regge
kinematics in the limit $s \to\infty$ and leading logarithmic
approximation. In the next-to-leading logarithmic approximation
(NLLA), corrections require also the contribution from the
quasi-multi-Regge kinematics~\cite{FL98}. Hence, the subenergies
between neighboring $s$-channel gluons must be large enough to be
in the Regge domain. At finite total energies, this implies that
the amplitude is represented by a finite sum of $N$
terms~\cite{PR}, where $N$ increases like $\ln s$, rather than by
the solution of the BFKL integral equation~\cite{FKL,BL,L}. The
interest in the first few terms of the series is related to the
fact that the energies reached by the present accelerators are not
high enough to accommodate a large number of $s$-channel gluons
that eventually hadronize and give rise to clusters of secondary
particles \cite{Lia}.

The lowest order diagram is that of two-gluon exchange, first
considered by Low and Nussinov~\cite{LN}. The next order,
involving an $s$-channel gluon rung was studied e.g. in
papers~\cite{BL,McCoy} and generalized in Ref. \cite{FKL}. The
problem of calculating these diagrams is twofold. The first one is
connected with the nonperturbative contributions to the scattering
amplitude in the ''soft'' region. It may be ignored by
''freezing'' the running coupling constant at some fixed value of
the momenta transferred and assuming that the forward amplitude
can be cast by a smooth interpolation to $t=0$. More consistently,
one introduces a nonperturbative model~\cite{Francesco} of the
gluon propagator valid also in the forward direction. The second
problem is more technical: as $s\to \infty $ the number of Feynman
diagrams that contribute to the leading order rapidly increases
and, in each of them, only the leading contribution is usually
evaluated. At any order in the coupling, sub-leading terms coming
both from the neglected diagrams and from the calculated ones are
present. Although functionally the result is always the sum of
increasing powers of logarithms, the numerical values of the
coefficients entering the sum is lost unless all diagrams are
calculated.

Conversely, one can expand the "supercritical" Pomeron $\sim
s^{\alpha_p(0)}$ in powers of $\ln(s)$. Such an expansion is
legitimate within the range of active accelerators, i.e. near and
below the TeV energy region, where fits to total cross sections by
 power or logarithms are known~\cite{Ezhela} to be equivalent
numerically. Moreover, forward scattering data (total cross
sections and the ratio of the real to the imaginary part of the
forward scattering amplitude) do not discriminate even between a
single and quadratic fit in $\ln(s)$ to the data.

In Ref. \cite{PR} a model for the Pomeron at $t=0$
based on the idea of a finite sum of ladder diagrams in QCD was
suggested. According to the idea of that paper, the number of
$s$-channel gluon rungs and correspondingly the powers of
logarithms in the forward scattering amplitude depends on the
phase space (energy) available, i.e. as energy increases,
progressively new prongs with additional gluon rungs in the
$s$-channel open. Explicit expressions for the total cross section
involving two and three rungs or, alternatively, three and four
prongs (with $\ln^2(s)$ and $\ln^3(s)$ as highest terms,
respectively) were fitted to the proton-proton and
proton-antiproton total cross section data in the accelerator
region.

In a related paper~\cite{bea} the Pomeron was considered as a
finite series of ladder diagrams, including one gluon rung besides
the Low-Nussinov "Born term" and resulting in a constant plus
logarithmic term in the total cross section. With a sub-leading
Regge term added, good fits to $pp$ and $p\bar p$ total as well as
differential cross section were obtained in~\cite{bea}. There is
however a substantial difference between Ref. \cite{PR} and that of
Ref.~\cite{bea} or simple decomposition in powers of $\ln(s)$,
namely that we consider the opening channels (in $s$) as threshold
effects, the relevant prongs being separated in rapidity by $\ln
s_0$, $s_0$ being a parameter related to the average subenergy in
the ladder. Although such an approach inevitably introduces new
parameters, we consider it more adequate in the framework of the
finite-ladder approach. We mention these attempts only for the
sake of completeness, although we stick to the simplest case of
$t=0$, where there are hopes to have some connection with the QCD
calculations.

Within the "finite gluon ladder approach" to the pomeron (see
\cite{PR} and references therein), several options are possible.
In Ref. \cite{PR} a system of interconnected equations was solved
with several free parameters, including the value of $s^i_0$. that
determine the opening of each threshold (prong). In that paper
finite gluon ladders were calculated from QCD, where the important
dynamical information is contained in $\rho$ of Eq. (7) of that
paper, including $\ln s$ terms multiplied by the QCD running
constant $\alpha_s$ constraining the interconnection between
various powers of the logarithms in the total cross section. If
one chooses $\alpha_s=0.5-0.7,$ a typical "frozen" value of the
QCD coupling constant, the resulting total cross section will rise
too fast with respect to the data. Good fits within this option
can be achieved only if  $\alpha_s$ is an order of magnitude
smaller than the above "canonical"' value. Whether this is
acceptable or not is an open question (see below, Sec. 4 and the
Conclusions of the present paper).

In the present paper we include the unitarization procedure: we
consider the QCD-inspired amplitude as a  Born term, subject to a
subsequent unitarization procedure. We use the eikonal formalism
and treat the running constant as a free parameter. The resulting
eikonalized amplitude, fitted to the data (Section 4), gives
$\alpha_s\approx0.2.$ This can be considered also as a way of
deriving the QCD running coupling in the soft region.

\section{Total cross sections from a finite sum of gluon ladders}

The Pomeron contribution to the total cross section is represented
in the form
\begin{equation}
\sigma_t^{(P)}(s)=\sum_{i=0}^N
f_i\:\theta(s-s_0^i)\:\theta(s_0^{i+1}-s)\;, \label{z1}
\end{equation}
where
\begin{equation}
f_i=\sum^i_{j=0}a_{ij}L^j\;, \label{z2}
\end{equation}
$s_0$ is the prong threshold, $\theta(x)$ is the step function and
$L\equiv \ln(s)$. Here, by $s$ and $s_0$, respectively,
$s$/(1\,GeV$^2$) and $s_0$/(1\,GeV$^2$) are implied. The main
assumption in Eq.~(\ref{z1}) is that the widths of the rapidity
gaps $\ln(s_0)$ are the same along the ladder. The functions $f_i$
are polynomials in $L$ of degree $i$, corresponding to finite
gluon ladder diagrams in QCD, where each power of the logarithm
collects all the relevant diagrams. When $s$ increases and reaches
a new threshold, a new prong opens adding a new power in $L$. In
the energy region between two neighbouring thresholds, the
corresponding $f_i$, given in Eq.~(\ref{z1}), is supposed to
represent adequately the total cross section.

In Eq.~(\ref{z1}) the sum over $N$ is a finite one, since $N$ is
proportional to $\ln(s)$, where $s$ is the present squared c.m.
energy. Hence this model is quite different from the usual
approach where, in the limit $s \to \infty$, the infinite sum of
the leading logarithmic contributions gives rise to an integral
equation for the amplitude.

To make the idea clearer, we describe the mechanism in the case of
three gaps (two rungs). To remedy the effect of the first
threshold and get a smooth behavior at low energies, we have
included also a Pomeron daughter, going like $\sim 1/s$ in the
first two gaps with parameters $b_0$ and $b_1$ respectively. Then
\begin{eqnarray}
f_0(s)  = a_{00}+b_0/s  && \;\;\;\mbox{for}\qquad s \leq s_0\;, \\
f_1(s)  = a_{10}+b_1/s+a_{11}L  && \;\;\;\mbox{for}\qquad s_0\leq
s \leq
s_0^2\;,\\
f_2(s)  = a_{20}+a_{21}L+a_{22}L^2  &&  \;\;\;\mbox{for}\qquad
s_0^2\leq s \leq s_0^3\;.
\end{eqnarray}
By imposing the requirement of continuity (of the cross section
and of its first derivative) one constrains the parameters. E.g.,
from the conditions $f_1(s_0)=f_0(s_0)$ and $f'_1(s_0)=f'_0(s_0)$
the relations
\begin{displaymath}
b_1=a_{11} s_0+b_0\;,
\end{displaymath}
\begin{displaymath}
a_{10}=a_{00}-a_{11}\ln (s_0)-a_{11}
\end{displaymath}
follow. Furthermore, from $f_2(s_0^2)=f_1(s_0^2)$ and
$f'_2(s_0^2)=f'_1(s_0^2)$ one gets
\begin{displaymath}
a_{20}=a_{22} \ln^2(s_0^2)+a_{10}+b_1 (1+\ln(s_0^2))/s_0^2\;,
\end{displaymath}
\begin{displaymath}
a_{21}=a_{11}-2a_{22}\ln (s_0^2)-b_1/s_0^2\;.
\end{displaymath}
The same procedure can be repeated for any number of gaps.

In fitting the model to the data, the authors of Ref. \cite{PR}
relied mainly on $p\bar p$ data that extend to the highest
(accelerator) energies, to which the Pomeron is particularly
sensitive. To increase the confidence level, $pp$ data were
included in the fit as well. To keep the number of the free
parameters as small as possible and following the successful
phenomenological approach of Donnachie and Landshoff~\cite{DL}, a
single ''effective'' Reggeon trajectory with intercept $\alpha
\left( 0\right) $ will account for nonleading contributions, thus
leading to the following form for the total cross section:
\begin{equation}
\sigma _t(s)=\sigma_t ^{(P)}(s)+R(s)\;,
\end{equation}
where $\sigma_t^{(P)}(s)$ is given by Eq.~(\ref{z1}) and $R(s)=a
{s}^{\alpha (0)-1}$ (note that $a$ is
different for $p\bar p$ and $pp$ and is considered as an
additional free parameter).

Ideally, one would let free the width of the gap $s_0$ and
consequently the number of gluon rungs (highest power of $L$).
Although possible, technically this is very difficult. Therefore
we considered only the cases of two and three rungs and, for each
of them, we treated $s_0$ as a free parameter.

Notice that the values of the parameters depend on the energy
range of the fitting procedure. For example, the values of the
parameters in $f_0$ if fitted in "its" range, i.e. for $s\leq
s_0$, will get modified in $f_1$ with the higher energy data and
correspondingly higher order diagrams included.

As a first attempt, only three rapidity gaps, that correspond to
two gluon rungs in the ladder were considered. Fits to the $p\bar
p$ and $pp$ data were performed from $\sqrt{s}=4$ GeV up to the
highest energy Tevatron data~\cite{abe}.  Interestingly, the value
of $s_0$ turned out to be very close to 144 GeV$^2$, i.e. the
value for which the energy range considered is covered with equal
rapidity gaps uniformly.

Next, the energy span available in the accelerator
region by four gaps, resulting in 3 gluon rungs and consequently
$L^3$ as the maximal power were covered \cite{PR}. After the matching procedure,
ten free parameters remained: first of all $s_0$, then
$a_{00},b_0,a_{11},a_{22},a_{32}$, $a_{33}$, each determined in
its range, while the two $a$'s and $\alpha \left( 0\right)$ are
fitted in the whole range of the data. The final value for $s_0$
turned out to be $s_0\simeq 42.5$ GeV$^2$ resulting in a sequence
of energy intervals ending at $\sqrt{s}=1800$ GeV. Interestingly,
search for the phase space region where the
production amplitude in the multicluster configuration has a
maximum resulted, with the help of cosmic ray data, to an average
"subenergy" $<s_i> \sim 44$ GeV$^2$~\cite{BP}, that is very near
to the value of $s_0$ found in the fit.

\section{Explicit iterations of BFKL}

From the theoretical point of view, the phenomenological model of
Section 2 corresponds to the explicit evaluation in QCD of gluonic
ladders with an increasing number of $s$-channel gluons. This
correspondence is far from literal since each term of the BFKL
series takes into account only the dominant logarithm in the limit
$s\to \infty$. In the following we give concrete expressions for
the forward high energy scattering amplitudes for hadrons in the
form of an expansion in powers of large logarithms in the leading
logarithmic approximation.

We start from known results obtained in paper~\cite{BL} where an
explicit expression for the total cross section for hadron-hadron
scattering has been obtained. In the high energy limit, it is
convenient to introduce the Mellin transform of the amplitude
\begin{displaymath}
{\cal A}(\omega,t)=\int_0^\infty d \tilde s
\tilde s^{-\omega-1}\frac{\mbox{Im}_s {\cal A}(s,t)}{s},\
\tilde s=\frac{s}{m^2}
\end{displaymath}
and its inverse
\begin{displaymath}
\frac{\mbox{Im}_s {\cal A}(s,t)}{s}=\frac{1}{2\pi i}
\int_{\delta-i\infty}^{\delta+i\infty} d\omega \tilde s^\omega
{\cal A}(\omega,t)\;.
\end{displaymath}
The general expression of $A(\omega,t)$ in the leading logarithmic
approximation has the form:
\begin{displaymath}
{\cal A}(\omega,t)=\int d^2k \frac{\Phi^a(k,q) \:
F_\omega^b(k,q)}{k^2(q-k)^2}\;,
\end{displaymath}
where $\Phi^a(k,q)$ and  $\Phi^b(k,q)$ (see next equation) are the
impact factors of the colliding hadrons $a$ and  $b$, obeying the
gauge conditions $\Phi^j(0,q)=\Phi^j(q,q)=0$ $(j=a,b)$. The
quantity $F_\omega^b(k,q)$ obeys the BFKL equation:
\begin{displaymath}
\omega F_\omega^b(k,q)=\Phi^b(k,q)+\gamma \int \frac{d^2k'}{2\pi}
\frac{A(k,k',q)F_\omega^b(k',q)-B(k,k',q)F_\omega^b(k,q)}{(k-k')^2}\;,
\end{displaymath}
with
\begin{displaymath}
A(k,k',q)=\frac{-q^2(k-k')^2+k^2(q-k')^2+k^{'2}(q-k)^2}{k^{'2}(q-k')^2}\;,
\end{displaymath}
\begin{displaymath}
B(k,k',q)=\frac{k^2}{k^{'2}+(k'-k)^2}+\frac{(q-k)^2}{(q-k')^2+(k-k')^2}\;,
\end{displaymath}
and
\begin{displaymath}
\gamma=3 \frac{\alpha_s}{\pi}\;.
\end{displaymath}
The strong coupling $\alpha_s$ is assumed to be frozen at a
suitable scale set, for example, by the external particles. The
iteration procedure and the inverse Mellin transform give
(furthermore, we set $q=0$):
\begin{displaymath}
\sigma_t(s)=\frac{\mbox{Im}_s A(s,0)}{s}=\int d^2k
\frac{\Phi^a(k,0)}{(k^2)^2}\: \left[\Phi^b_0(k)+\rho
\Phi_1^b(k)+\frac{1}{2!}\rho^2\Phi^b_2+...\right]\;,
\end{displaymath}
where
\begin{equation}
\rho=\frac{3\alpha_s}{\pi}\ln {\tilde s}
\label{zr}
\end{equation}
and the subsequent iterations begin from
$\Phi_0^b(k)=\Phi^b(k,0)$. In the previous integral and everywhere
in the following, all the momenta are 2-dimensional Euclidean
vectors, living in the plane transverse to the one formed by the
momenta of the colliding particles.

To obtain the cross section of proton-proton scattering, we use
the ansatz of Ref.~\cite{levin} for the impact factor of a hadron
in terms of its form factor $F(q^2)$:
\begin{displaymath}
\Phi^p(k,q)=F^p\left(\frac{q^2}{4}\right)
-F^p\left(\left(k-\frac{q}{2}\right)^2\right)\;,\;\;\;
\Phi^p(0,q)=\Phi(q,q)=0\;.
\end{displaymath}
Here the 2-dimensional Euclidean vector $q$ is related to the
4-dimensional transferred momentum $Q$ by the relation
$Q^2=-q^2<0$. To get the input value for $\Phi_0$, we use
\begin{equation}
F_0(k)=a k^2 e^{-ck^2}\;, \label{zf}
\end{equation}
where $a$ and $c$ are in GeV$^{-2}$. It is convenient to define
\begin{displaymath}
\psi_n(k^2)=\frac{F_n(k)}{k^2}\;,
\end{displaymath}
then (for $n\geq 1$)
\begin{displaymath}
\psi_n(k^2)=\int_0^1\frac{dx}{1-x}\biggl(\psi_{n-1}(k^2x)-\psi_{n-1}(k^2)\biggr)
+\int_1^\infty
\frac{dx}{x-1}\left(\psi_{n-1}(k^2x)-\frac{1}{x}\psi_{n-1}(k^2)\right)
\end{displaymath}
and
\begin{equation}
\sigma_t(s)=\pi\int_0^\infty dk^2
\psi_0(k^2)\sum_n\psi_n(k^2)\frac{\rho^n}{n!}\;. \label{zz}
\end{equation}
Integrations can be performed analytically, due to the simple
choice of the impact factor in Eq.~(\ref{zf}), and the final
result is:
\begin{displaymath}
\sigma_t(s)=\frac{\pi a^2}{2c} \left\{ 1+2 (\ln
2)\rho+\left[\frac{\pi^2}{12}+ 2 (\ln 2)^2 \right]\rho^2+ \right.
\end{displaymath}
\begin{equation}
\left. \frac{1}{3} \left[\frac{\pi^2}{2}(\ln 2)+4 (\ln
2)^3-\frac{3}{4}\zeta(3)\right] \rho^3+\ldots \right\}\;.
\label{za}
\end{equation}
where $\rho$ is defined in Eq.~(\ref{zr}) and $\zeta(3)$ is the Riemann's Zeta function
$\\ \zeta(3)\approx1.202.$

\section{Unitarized fit}

In section II we quoted the calculated cross sections of the
finite-laddear-Pomeron model with the parameters fitted to the
existing data, the threshold value (opening prong) of the
interconnected ladders playing there an important role. In Sec. 3
the parameters were calculated from QCD. With these parameters the
resulting cross sections overshoot the data, which is not
surprising, since the calculated Born term should be subjected to
a unitarization procedure. Below we perform such calculations in
the framework of the eikonal formalism and compare the results
with the experimental data.

We start from Eq. (10), Sec. III for the $pp$ and $\bar pp$ total
cross section. Supplying that expressing with an exponential
$t$-dependence we get the elastic scattering amplitude:
\begin{equation}\label{amplitude}
F_{Born}(s,t)=A(-i
\tilde s)^{1+\alpha't}[a_0+a_1\gamma\ln(-i \tilde s)+a_2\gamma^2\ln^2(-i
\tilde s)+a_3\gamma^3ln^3(-i \tilde s)]e^{Bt},
\end{equation}
where $\alpha'$ and $B$ are new fitting parameters, and

$$a_0=1+ \frac{\pi^2}{4}\Bigl(\frac{\pi^2}{12} + 2\ln^22\Bigr) \gamma^2, $$

 $$a_1=\frac{\pi^2}{4} \Bigl[\frac{\pi^2}{2}\ln2+ 4\ln ^32- \frac{3}{4}\zeta(3)\Bigr]\gamma^2 +2 \ln2,$$
$$a_2=\frac{\pi^2}{12}+2\ln^22,$$
$$ a_3=\frac{1}{3}\Bigl[\frac{\pi^2}{2}\ln 2+4\ln^32-\frac{3}{4}\zeta(3)\Bigr],$$

$$A=-\frac{a^2}{8c},$$
and is the Riemann $\zeta$-function, defined above.

In the eikonalization procedure we follow Ref. \cite{Kroll},
according to which the Pomeron amplitude
\begin{equation}\label{Kroll}
F_P(s,t)=is\int_0^\infty
bdbJ_0\Bigl(b\sqrt{-t}\Bigr)\Bigl(1-e^{i\chi(b,s)}\Bigr),
\end{equation}
where $J_0$ is the Bessel function of zeroth order
 and the
eikonal $\chi$ is
\begin{equation}\label{eikonal}
\chi(s,b)=\frac{1}{s}\int_0^{\infty}\sqrt{-t}d\sqrt{-t}I_0(b\sqrt{-t})F_{Born}(s,t).
\end{equation}

Inserting the expression for the Pomeron into Eq. (\ref{eikonal}) and expanding the exponential in
(\ref{Kroll}), one find  for the eikonalized Pomeron amplitude



\begin{equation}\label{series}F_P=2is\xi \sum_{k=1}^\infty\frac{1}{kk!}
 \left(-\frac{\xi}{\mu}\right)^{k-1}e^{\mu t/k}.\end{equation}

\begin{figure}[h]
\begin{center}
\hspace{-1.cm}
\includegraphics[width=0.8\textwidth,angle=0]{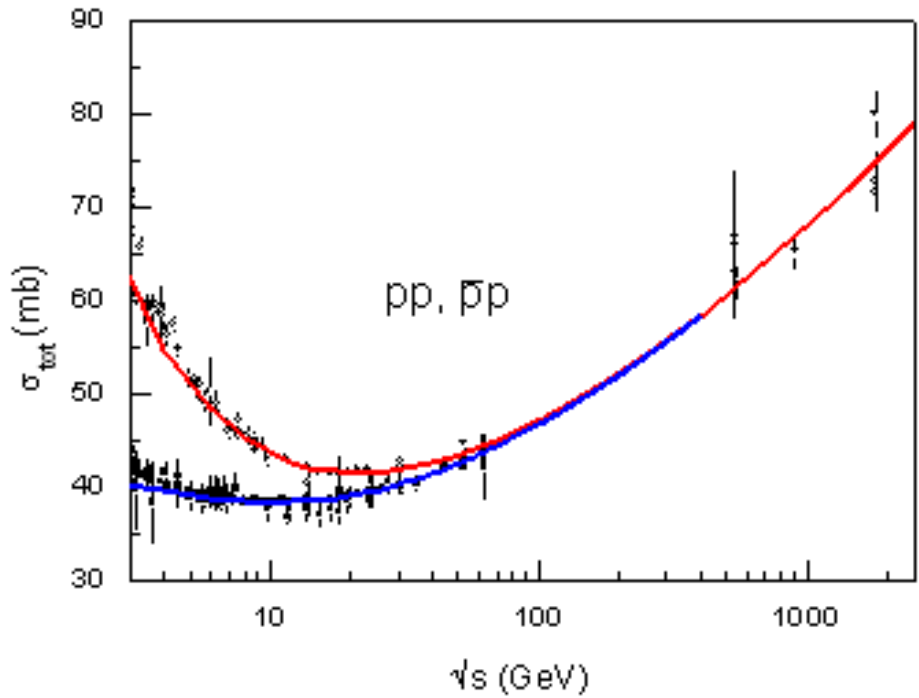}
\caption{\small \it {Total $pp$ (lower, blue line) and $\bar pp$
(upper, red line) cross sections from the uniratized (eikonalized)
version of the model.}} \label{fig:fit_new}
\end{center}
\end{figure}

\begin{figure}[h]
\begin{center}
\hspace{-1.cm}
\includegraphics[width=0.8\textwidth,angle=0]{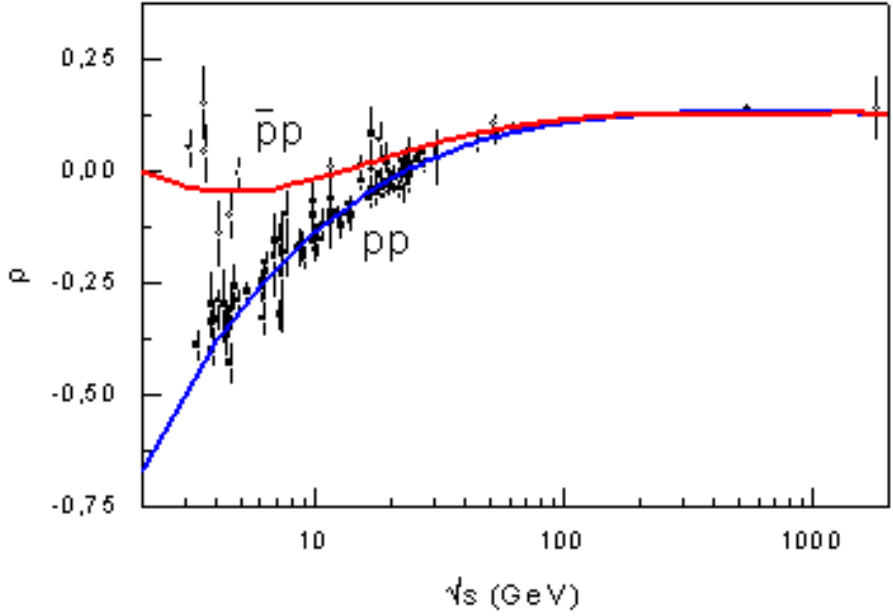}
\caption{\small \it {The ratio $\rho(s)=Re A(s,0)/Im A(s,0)$ from
the same model.}} \label{fig:fit_new}
\end{center}
\end{figure}

Respectively the forward Pomeron
amplitude is
\begin{equation}\label{result}
F_P(s,t=0)=2is\mu[C+ln(\xi/\mu)+E_1(\xi/\mu)],
\end{equation}
where  \begin{equation}\mu=B+\alpha'ln(-i \tilde s);\  \end{equation}
$\xi=\frac{A}{2m^2}(\xi_0+\xi_1+\xi_2+\xi_3),\ \
$ and
$$\xi_0=a_0,\ \
\xi_1=a_1\gamma\ln(-i\tilde s),$$
$$\xi_2=a_2\gamma^2\ln^2(-i\tilde s),\ \
\xi_3=a_3\gamma^3\ln^3(-i\tilde s),$$ $C=0.577216 $ is the Euler
constant and $E_1$ is the asymptotic form of the first order
exponential integral:

\begin{equation}E_1=\frac{\exp({\xi/\mu})}{\xi/\mu} [1-\frac{1}{\xi/\mu}+
\frac{2}{(\xi/\mu)^2}-
\frac{6}{(\xi/\mu)^3}
 +....].
 \end{equation}

The obtained eikonalized Pomeron term is appended by a
contributions from secondary reggeons, $f$ and $\omega$:
$$F_R^{\pm}(s,t=0)=g_f\tilde s^{\alpha_f(0)}\pm
ig_{\omega}\tilde s^{\alpha_\omega(0)},$$ where the $+(-)$ sign
corresponds to $\bar pp(pp)$ scattering, the resulting forward
amplitude being
$$F^{\bar pp}_{pp}(s,t=0)=F_P(s,t=0)+ F_R(s,t=0).$$
For the total cross section the norm
$$\sigma=\frac{4\pi}{s}Im F^{\bar pp}_{pp}(s,t=0)$$
was used and $\rho(s,0)=Re F^{\bar pp}_{pp}(s,t=0)/Im F^{\bar pp}_{pp}(s,t=0).$

Fits of the total of 8 free parameters to $238$ data points on $pp$ and $\bar pp$ total cross-sections
as well as on the ratio $\rho$   (see Table 1) were performed in the range $5$ GeV $\leq\sqrt{s}\leq 1.8$ TeV
with the results shown in Figs. 1 and 2 and Table 2.  Predictions for $pp$ at the expected LHC energies are
quoted in Table. 3.

The quality of the fits ($\chi^2/dof)$ is good, comparable to that
in e.g. Ref. \cite{Cudell} (for a recent review on the subject see
Ref. \cite{LHC}).

\begin{table}
\caption{Number of used data points and $\chi^{2}$ per degree of
freedom ($\chi^{2}$/dof) for 5 GeV $\leq$ $\sqrt{s}$
$\leq$ 1.8 TeV (same as  in \cite{Cudell})}\vskip3mm
\begin{center}{\small \begin{tabular}{|lc|} \hline
  $\sigma_{pp}$ & 104  \\
  \hline
  $\sigma_{\overline{p}p}$ &  59 \\
  \hline
  $\rho_{pp}$ & 64  \\
  \hline
  $\rho_{\overline{p}p}$ &  11 \\
  \hline
  Total number of points & 238  \\
  \hline
Number of free parameters & 8 \\
\hline
$\chi^2/$dof & 1.11 \\
\hline
\end{tabular}}
\end{center}
\end{table}

\begin{table}
\noindent\caption{Values of the fitted parameters}\vskip3mm
\begin{center}{\small
\begin{tabular}{|l|r|l|}
\hline
Parameter  & Value & error\\
\hline
  $A$ &  0.526 &  0.198\\
\hline
  $\alpha_{s}$ & 0.190 & 0.033\\
\hline
  $B$ &  0.116 & 0.121\\
\hline
$\alpha'$ & 0.134 & 0.004  \\
\hline
  $g_{f}$ &  -4.56 &0.35\\
\hline
  $\alpha_{f}(0)$ &  0.858 & 0.086\\
\hline

$g_{\omega}$ &  3.73 & 0.16\\
\hline
  $\alpha_{\omega}(0)$ &  0.451 & 0.013\\
\hline
\end{tabular}}
\end{center}
\end{table}

\begin{table}
\noindent\caption{Prediction of the model for the LHC
energies}\vskip3mm
\begin{center}{\small\begin{tabular}{|l|r|l|} \hline
energy (TeV)&  6  & 12 \\
\hline
  $\sigma(pp$$/\overline{p}p)(mb)$ & 91.3& 101  \\

\hline
  $\rho(pp$$/\overline{p}p)$ & 0.121& 0.115  \\

  \hline
\end{tabular}}
\end{center}
\end{table}

\section{Conclusions}

Our main goal was an adequate picture of the Pomeron exchange at
$t=0$.  In our opinion, it is neither an infinite sum of gluon
ladders as in the BFKL approach~\cite{FKL,BL,L}, nor its power
expansion. In fact, the finite series - call it "threshold
approach" - considered in Sec. II and in the previous
papers~\cite{PR} realizes a non-trivial dynamical balance between
the total reaction energy and the subenergies equally partitioned
between the multiperipheral ladders.

The role and the value of the width of the gap, $s_0$, is an
important physical parameter {\sl per se}, independent of the
model presented above. We have fitted it and compared successfully
with the prediction from cosmic-ray data. However its value may be
estimated e.g. as the lowest energy where the Pomeron exchange is
manifest, although the latter is also a matter of debate.

The case of two terms (logarithmic rise in $s$)
is particularly interesting as it corresponds to a dipole Pomeron
with a number of attractive features~\cite{JShS} such as
self-reproducibility with respect of unitarity corrections. In
case of a $\ln^2(s)$ rise (three terms) we still should not worry
about the Froissart bound, so ultimately the Pomeron as viewed in
this paper does not need to be unitarized. This conclusion is an
important by-product of our paper. For the dipole Pomeron,
relevant calculations for $t\neq 0$ are interesting and important
but difficult. In the case of a single gluon rung they were
performed in Ref.~\cite{bea} and, with a non-perturbative gluon
propagator, in the last reference of~\cite{Francesco}.
It is significant that the obtained in this paper value of the $\alpha_s$
as fitted parameter corresponds to one calculated with describing $F_2$ through
of finite sum of gluon ladders ~\cite{machado}, which is typical of this kinematical region.

As mentioned in the Introduction, acceptable high-energy
asymptotics with the QCD Pomeron of Sec. III, without
unitarization, can be achieved only at the expense of
substantially lowering the canonical, "frozen" QCD running
constant. The unitarization procedure and the subsequent fit to
the data, presented in Sec. 4 give independently the value
$\alpha_s=0.19,$ thus providing one more means for the
determination of this fundamental constant of QCD.

\section*{Acknowledgments}

We thank Francesco Paccanoni and Alessandro Papa for discussions
and for their collaboration at an earlier stage of this work. We
also acknowledge discussions and useful remarks by Victor
Abramovsky and Natalia Radchenko. L.J. appreciates the hospitality
and support of the University of Calabria and INFN, where this
work was completed. R.F. acknowledge partial support by the
Italian Ministry of University and Research.

\end{document}